\documentclass[prl,twocolumn,superscriptaddress]{revtex4-1}
\usepackage{amssymb}
\usepackage{amsfonts}
\usepackage{bbm}
\usepackage{mathrsfs}
\usepackage{graphics,graphicx,epsfig,bm,amsmath,amsthm,amssymb}
\usepackage{bm}
\usepackage{bbm}
\usepackage{longtable}
\usepackage{multirow}
\usepackage{array}
\usepackage{color}
\usepackage{cases}
\usepackage{booktabs }
\usepackage[usenames,dvipsnames]{xcolor}
\newcommand{\ket}[1]{\left| #1\right\rangle}
\usepackage{xcolor}

\usepackage{float}
\usepackage[a4paper,colorlinks=true,
linkcolor=blue,citecolor=blue,
pdfauthor={ },
pdftitle={ },
pdfsubject={ },
pdfkeywords={ }]{hyperref}

\begin{document}

\title{Constraints on a Spin-Dependent Exotic Interaction between Electrons with Single Electron Spin Quantum Sensors}

\author{Xing Rong}
\affiliation{CAS Key Laboratory of Microscale Magnetic Resonance and Department of Modern Physics, University of Science and Technology of China, Hefei 230026, China}
\affiliation{Synergetic Innovation Center of Quantum Information and Quantum Physics, University of Science and Technology of China, Hefei 230026, China}
\affiliation{Hefei National Laboratory for Physical Sciences at the Microscale, University of Science and Technology of China, Hefei 230026, China}

\author{Man Jiao}
\affiliation{CAS Key Laboratory of Microscale Magnetic Resonance and Department of Modern Physics, University of Science and Technology of China, Hefei 230026, China}
\affiliation{Synergetic Innovation Center of Quantum Information and Quantum Physics, University of Science and Technology of China, Hefei 230026, China}
\affiliation{Hefei National Laboratory for Physical Sciences at the Microscale, University of Science and Technology of China, Hefei 230026, China}

\author{Jianpei Geng}
\affiliation{CAS Key Laboratory of Microscale Magnetic Resonance and Department of Modern Physics, University of Science and Technology of China, Hefei 230026, China}
\affiliation{School of Electronic Science and Applied Physics, Hefei University of Technology, Hefei 230009, China}

\author{Bo Zhang}
\email{bz8810@ustc.edu.cn}
\affiliation{CAS Key Laboratory of Microscale Magnetic Resonance and Department of Modern Physics, University of Science and Technology of China, Hefei 230026, China}
\affiliation{Synergetic Innovation Center of Quantum Information and Quantum Physics, University of Science and Technology of China, Hefei 230026, China}
\affiliation{Hefei National Laboratory for Physical Sciences at the Microscale, University of Science and Technology of China, Hefei 230026, China}

\author{Tianyu Xie}
\affiliation{CAS Key Laboratory of Microscale Magnetic Resonance and Department of Modern Physics, University of Science and Technology of China, Hefei 230026, China}
\affiliation{Synergetic Innovation Center of Quantum Information and Quantum Physics, University of Science and Technology of China, Hefei 230026, China}
\affiliation{Hefei National Laboratory for Physical Sciences at the Microscale, University of Science and Technology of China, Hefei 230026, China}

\author{Fazhan Shi}
\affiliation{CAS Key Laboratory of Microscale Magnetic Resonance and Department of Modern Physics, University of Science and Technology of China, Hefei 230026, China}
\affiliation{Synergetic Innovation Center of Quantum Information and Quantum Physics, University of Science and Technology of China, Hefei 230026, China}
\affiliation{Hefei National Laboratory for Physical Sciences at the Microscale, University of Science and Technology of China, Hefei 230026, China}

\author{Chang-Kui Duan}
\affiliation{CAS Key Laboratory of Microscale Magnetic Resonance and Department of Modern Physics, University of Science and Technology of China, Hefei 230026, China}
\affiliation{Synergetic Innovation Center of Quantum Information and Quantum Physics, University of Science and Technology of China, Hefei 230026, China}
\affiliation{Hefei National Laboratory for Physical Sciences at the Microscale, University of Science and Technology of China, Hefei 230026, China}

\author{Yi-Fu Cai}
\affiliation{CAS Key Laboratory for Research in Galaxies and Cosmology, Department of Astronomy, University of Science and Technology of China, Hefei 230026, China}
\affiliation{School of Astronomy and Space Science, University of Science and Technology of China, Hefei 230026, China}
\author{Jiangfeng Du}
\email{djf@ustc.edu.cn}
\affiliation{CAS Key Laboratory of Microscale Magnetic Resonance and Department of Modern Physics, University of Science and Technology of China, Hefei 230026, China}
\affiliation{Synergetic Innovation Center of Quantum Information and Quantum Physics, University of Science and Technology of China, Hefei 230026, China}
\affiliation{Hefei National Laboratory for Physical Sciences at the Microscale, University of Science and Technology of China, Hefei 230026, China}

\date{\today}

\begin{abstract}
A new laboratory bound on the axial-vector mediated interaction between electron spins at micrometer scale  is established with single nitrogen-vacancy (NV)
centers in diamond. A single crystal of $p$-terphenyl doped pentacene-$d_{14}$ under laser pumping provides the source of polarized electron spins. Based on the measurement of polarization signal via nitrogen-vacancy centers, we set a constraint for the exotic electron-electron coupling $g_A^eg_A^e$, within the force range  from 10 to 900 $\mu$m. The obtained upper bound of the coupling at 500 $\mu$m is $|g_A^eg_A^e / 4\pi\hbar c |\leq 1.8\times 10^{-19} $, which is one order of magnitude more stringent than a previous experiment. Our result shows that the NV center can be a promising platform for searching for new particles predicted by theories beyond the standard model.
\end{abstract}


\maketitle

Given our ignorance of the ultraviolet completion of particle physics, it is of great importance to investigate new particles beyond the standard model \cite{RMP_ExtSM}.
Theoretical predicted particles, such as pseudoscalars fields (axion and axionlike particles \cite{AnnuRevNuclPartSci_ALP,PRD_NewMacroForce}) and axial-vector fields (paraphotons \cite{Dobrescu_PRL_2005} and extra $Z$ bosons \cite{Appelquist_PRD_2003, JHighEnergyPhys_MacroForce}), have stimulated attentions in a wide variety of researches \cite{Kimball}. It has been well motivated for decades from the requirement of cosmology \cite{PR_axion_cosmology}, namely, the candidates of dark matter \cite{PhysRep_DM} and dark energy \cite{PRL_DE}, and from the understanding of the symmetries of charge conjugation and parity in quantum chromodynamics (QCD) \cite{PRL_PQ}, as well as predictions from string theory \cite{RMP_ExtSM}.
The exchange of these hypothetical particles results in spin-dependent forces \cite{JHighEnergyPhys_MacroForce}, which were originally discussed by Moody and Wilczek \cite{PRD_NewMacroForce}.
Various laboratory searching experiments focus on the detection of macroscopic axial-vector dipole-dipole forces between polarized electrons, described by $V_2$ potentials in Ref. \cite{JHighEnergyPhys_MacroForce}, ranging from the atomic scale to the radius of the Earth \cite{Kimball}.
The series of stringent constraints on this coupling have been set by torsion-pendulum experiments \cite{Ritter_PRD_1990,Heckel_PRD_2008}, a trapped ions experiment \cite{PRL_ConstraintDipDip}, positronium hyperfine spectroscopy \cite{Karshenboim2010,Leslie2014,Kimball}, helium fine-structure spectroscopy \cite{Ficek2017}, and by using polarized electrons within the Earth \cite{Hunter928}. Recently, data from STM-ESR experiments \cite{STM_ESR_science,STM_ESR_Nat_Nano} have been used to constrain exotic dipole-dipole interactions between electrons at nanometer scale \cite{Luo_PRD}.

In this Letter, we established a new constraint on an exotic dipole-dipole interaction between electrons at the micrometer scale by single nitrogen-vacancy (NV) centers in diamond. The source of polarized electrons was provided by a single crystal of $p$-terphenyl doped pentacene-$d_{14}$ under laser pumping \cite{Sloop_1981}. The sensor can be engineered to be sensitive to the signal from polarized electrons\cite{Xie_2017}. Based on our recent measurement of polarized electrons by NV centers, we have established a new constraint on the axial-vector mediated interaction between electrons at micrometer scale, which considerably improves on previous experimental bounds.

\begin{figure}
\centering
\includegraphics[width=\columnwidth]{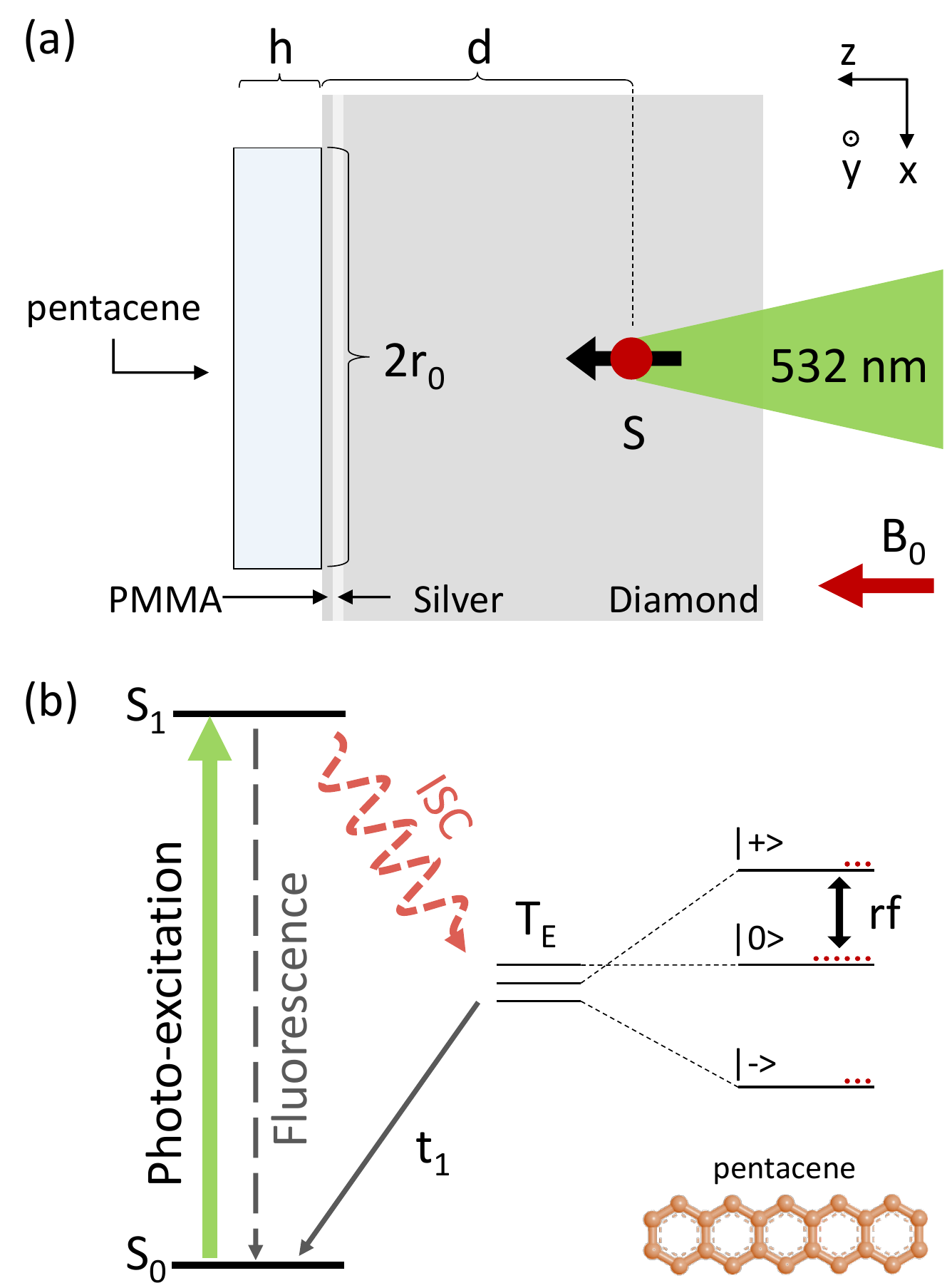}
\caption{(a) Schematic experimental setup. A NV center in diamond, which is labeled as $S$, is used to search for the exotic dipole-dipole interaction between electrons. The source of polarized electrons is provided by laser pumped pentacene doped in a $p$-terphenyl single crystal with the thickness being $ h = 15~\mu$m. The radius of the laser beam for pumping pentacene is $r_0 = 35~\mu$m. The distance between the NV center and the surface of the single crystal is labeled by d. An external magnetic field, $B_0 = 512~$G, was applied along the NV's axis (labeled by the $z$ axis).
(b) The electronic energy diagram of pentacene within a $p$-terphenyl host lattice under external magnetic field $B_0$. A laser with a 520-nm wavelength is used to pump the state of pentacene from the singlet ground state $S_0$  to the first excited single state $S_1$. Then, they decay quickly into the triplet state $T_E$ through spin-selective intersystem crossing (ISC). The population in the state $\ket{0}$ is much larger than that in $\ket{\pm}$. A radio-frequency pulse labeled by rf is to engineer the population of states $\ket{0}$ and $\ket{+}$. The spin polarization will relax back to the ground singlet state by phosphorescence or nonradiative decay. The decay time of states $\ket{\pm}$ is $t_1 = 7\pm1~\mu$s \cite{Xie_2017}.
\label{fig1}}
\end{figure}

Single NV centers in diamond, which are defects composed of a substitutional nitrogen atom and a neighboring vacancy\cite{Doherty2013}, have been proposed as quantum sensors for detecting a weak magnetic signal within the nanoscale \cite{NP_NV_2008,Degen_RMP2017}.
The size of this quantum sensor can be engineered  to be small compared to the micrometer force range, and the geometry enables close proximity between the sensor and the source. Furthermore, the magnetic noise can be isolated well by dynamical decoupling technology \cite{Du_2009,Degen_RMP2017}.  Recently, this type of quantum sensor has been proposed and utilized to explore electron-nucleon monopole-dipole interaction \cite{Rong_2018}.

\begin{figure}
\centering
\includegraphics[width=\columnwidth]{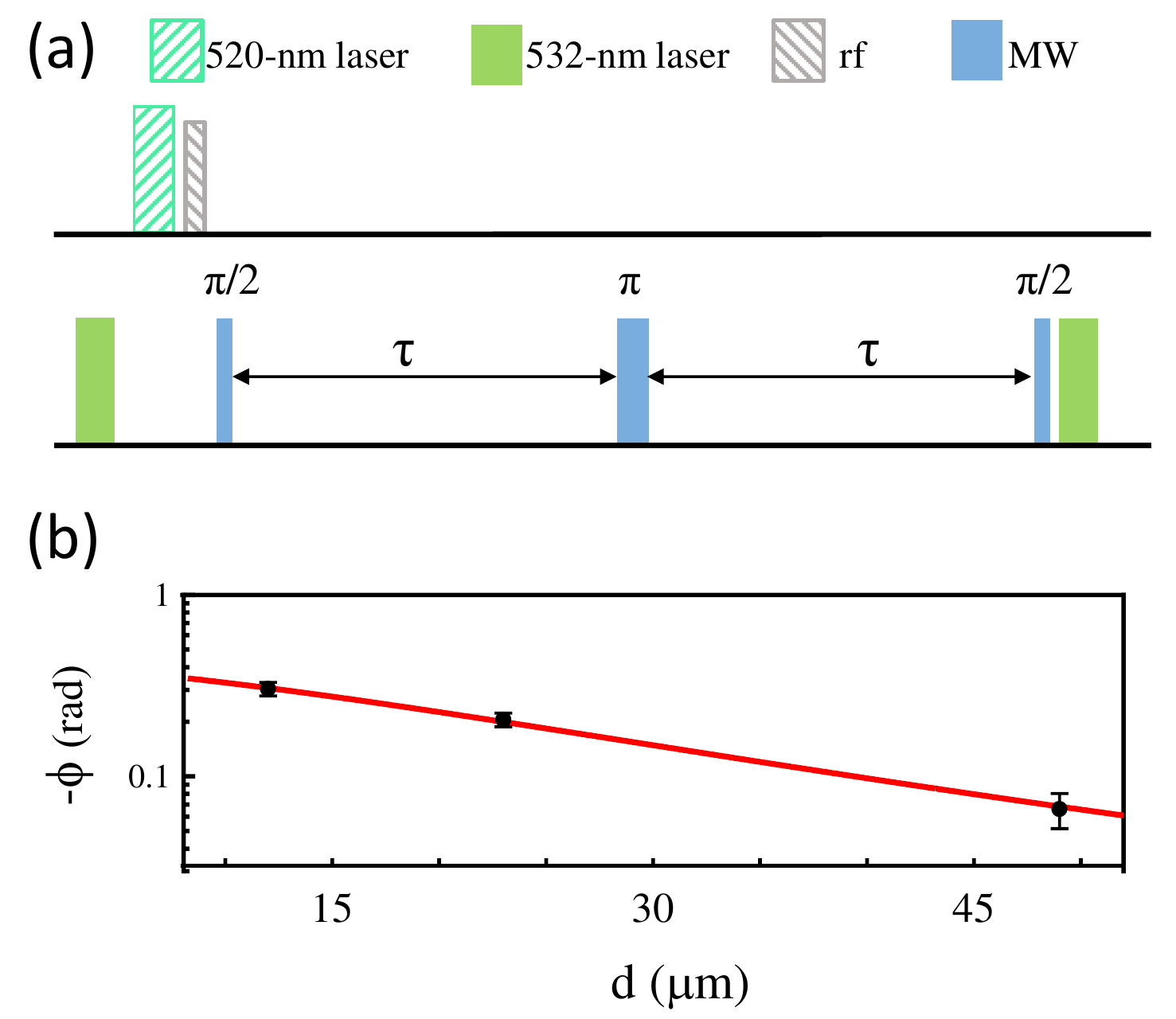}
\caption{(a) Pulse sequence for measurement of the polarized electrons by $S$. For the rf pulse,  the frequency is $820~$MHz, and the pulse length is 80 ns. For MW pulses, the frequency is $1.43~$GHz, and the pulse length of $\pi/2$($\pi$) is about 90 ns (180 ns). The pulse lengths of all the laser pulses are  1.5 $\mu$s. The delay time between MW pulses is set to $\tau = 30~\mu$s.
(b) Black circles with error bars are experimental accumulated $\phi$ due to the polarized electrons, with different distances $d$ = 12, 23, 49 $\mu$m. The error bars of the data are due to the photon statistics. Red line is the fit for $\phi$ with Eq. \ref{eqn4} when $\lambda = 500 ~\mu$m.
\label{fig2}}
\end{figure}

Herein, we focus on searching for exotic dipole-dipole interaction mediated by  axial-vector fields between electrons.
Figure \ref{fig1}(a) shows the schematic of the setup. A single crystal of $p$-terphenyl doped with pentacene-$d_{14}$, $0.05$ mol$\%$, is placed on the surface of the diamond.
The spin density of the sample is estimated to be $\rho = 1.62 \times 10^{-3}~$nm$^{-3}$.
The thickness of the single crystal is $h = 15~\mu$m.
The long axis of the pentacene molecule is nearly along the [111]-NV axis.
A 520-nm laser pulse with the beam intensity being about $10^7~$W/m$^2$ is applied on the single crystal to generate polarized electrons \cite{Sloop_1981}.
The NV center labeled by $S$ is a few micrometers below the surface of the diamond.
The ground state of the NV center is an electron spin triplet state $^3A_2$ with three substates $\ket{m_S=0}$ and $\ket{m_S=\pm1}$\cite{Doherty2013}.
A static magnetic field $B_0$ of about 512~G is applied along the NV symmetry axis to remove the degeneracy of the $|m_S=\pm1\rangle$ spin states.
The spin states $\ket{m_S=0}$ and $\ket{m_S=-1}$ are utilized as a quantum sensor \cite{Degen_RMP2017}.
The state of $S$ can be manipulated by microwave pulses labeled by MW in Fig. \ref{fig2}(a), which are delivered by a copper microwave wire. The $\ket{m_S = +1}$ state remains idle due to the large detuning.
Laser pulses with wavelengths of $532~$nm are applied to initialize and readout the state of $S$ \cite{Rong_2018}. There are two layers, a 150-nm-silver layer and a 100-nm-PMMA layer, between the single crystal and diamond to isolate  the two laser beams as well as fluorescence from S and the single crystal.

The first step is to prepare polarized electrons. The electronic energy level diagram of  electron spins in a pentacene molecule \cite{Ong_1993} is shown in Fig. \ref{fig1}(b).
After being excited by a 520-nm laser pulse, the pentacene can be pumped from the singlet state $S_0$ to the triplet manifold $T_E$  via spin selective intersystem crossing \cite{Patterson_1984,Koehler_1996}.
The population of the state $\ket{0}$ of the triplet sublevels is much greater than the states $\ket{\pm}$, while the populations of $\ket{\pm}$ are equal \cite{Sloop_1981}.
In our experiment, a $1.5~\mu$s laser pulse by a Gaussian beam with the radius of $35~\mu$m was applied.
A radio-frequency (rf) pulse with frequency resonant to the transition between $\ket{0}$ and $\ket{+}$ is applied after the laser pumping. The frequency of rf is set to 820~MHz, and the time duration of the rf pulse is $80~$ns.
After this rf pulse, a nonzero population difference  between the Zeeman eigenstates of an external magnetic field ($\ket{\pm 1}_p$), with $P_0 $ being about $0.5\%$, is generated \cite{Xie_2017}.
After the polarization procedure, the state of the electron spins will relax back to the singlet ground state $S_0$, which is of magnetic resonance
silence.
This results in a decay of polarization $P(t) = P_0 \exp(-t/t_1)$ with decay time $t_1 = 7 \pm1\mu$s.

Now, we consider the interactions between polarized electrons of  pentacene and $S$.
The magnetic diople-diople interaction between a single electron spin and $S$ is
\begin{equation}
H_1 =  - \frac{\mu_0\gamma_e\gamma_e\hbar^2}{16\pi r^3}[3(\vec{\sigma_1}\cdot\hat{r})(\vec{\sigma_2}\cdot\hat{r})- (\vec{\sigma_1}\cdot\vec{\sigma_2})],
\end{equation}
where $\vec{\sigma_1}$ and $\vec{\sigma_2}$ stand for Pauli vectors of the electron spin of pentacene and that of $S$, respectively, and $\gamma_e = 2\pi\times 2.8~$MHz/Gauss stands for the gyromagnetic ratio of the electron spin. The symbol $\vec{r}$ is the displacement vector between the electrons, and $r=|\vec{r}|$ and $\hat{r}=\vec{r}/r$ are the displacement and the unit displacement vector, respectively. The axial-vector dipole-dipole interaction mediated by hypothetical axial-vector bosons \cite{JHighEnergyPhys_MacroForce} can be written as
\begin{equation}
H_2 = \frac{g_A^eg_A^e}{4\pi\hbar c}\frac{\hbar c}{r} (\vec{\sigma_1}\cdot \vec{\sigma_2})e^{{-\frac{r}{\lambda}}},
\end{equation}
where $g_A^e g_A^e/4\pi\hbar c$ is the dimensionless axial-vector coupling constant between electrons, $\lambda=\hbar/(mc)$  is the force range, $m$ is the mass of the hypothetical particle, $\hbar$ is Plank's constant divided by $2\pi$, and $c$ is the speed of light.
When the electron spin of pentacene is in the state of $\ket{+1}_p$, the quantum sensor S feels an effective magnetic field from the electron spin, which can be written as
\begin{equation}
b_{\textrm{eff}}(r,\theta) = - \frac{\mu_0\gamma_e\hbar}{8\pi r^3}(3\cos^2 \theta - 1)  + (\frac{g_A^e g_A^e}{4\pi\hbar c})\frac{2c}{\gamma_e}\frac{e^{-\frac{r}{\lambda}}}{r},
\label{eq3}
\end{equation}
where $\theta$ stands for the angle between the external magnetic field and $\vec{r}$.
The first term in the right part of Eq. \ref{eq3} is due to the magnetic dipole-dipole interaction, and the second term is from the axial-vector coupling between electrons.
The effective magnetic field felt by $S$ from a bulk of pentacene with the electron spin density $\rho$ and polarization $P(t)$ is
\begin{equation}
b(t) = \rho P(t)\int_V b_{\textrm{eff}}(r,\theta) dV,
\label{eqn4}
\end{equation}
where $V$ stands for the cylinder of polarized electrons. The radius of the cylinder is equal to the radius of the laser beam, which is $35\pm 5~\mu$m. The thickness of the cylinder is $15\pm3 ~\mu$m.

The experimental pulse sequence is shown in Fig. \ref{fig2}(a).  The first $\pi/2$ microwave pulse prepares $S$ to a superposition state $(|0\rangle-i|1\rangle)/\sqrt{2}$. The spin echo sequence \cite{PR_Echo} has been applied on $S$ to cancel unwanted semistatic magnetic noise, and the coherence time of $S$ is about $400~\mu$s. The delay time $\tau$ in the pulse sequence is fixed to  be $\tau= 30~\mu$s, which is much shorter than $T_2$ and is much longer than the decay time of polarized electrons $t_1$.
A 520-nm laser pulse together with a rf pulse are applied on the bulk pentacene to prepare polarized electrons.
The polarized electrons generate an effective magnetic field $b(t)$ on the NV center's electron spin via the coupling between them.
This effective magnetic field $b(t)$ causes a phase shift $\phi = \int_0 ^\tau \gamma_e b(t) dt  - \int_\tau ^{2\tau} \gamma_e b(t) dt $ on the state of $S$ and the final $\pi/2$ pulse converts this phase shift to the population of the state $\ket{m_S=0}$ of $S$.
The phase of the last $\pi/2$ is set to be $90^\circ$ relative to the first $\pi/2$ pulse, so that the accumulated phase due to coupling to the polarized electrons can be obtained by $\phi = \arcsin (1- 2P_{\ket{m_S = 0}})$, where $P_{\ket{m_S = 0}}$ stands for the population in state $\ket{m_S = 0}$ of NV centers.

The experimental data are presented in Fig. \ref{fig2}(b).
Three NV centers with different depths have been chosen to measure the signal from the polarized electrons.
The three depths of these NV centers are measured to be $d$ = 12, 23, and 49$~\mu$m \cite{SupplementalMaterial}.
The experimental phases acquired by NV centers are presented as black circles with error bars in Fig. \ref{fig2}(b). The error bars are due to the photon statistics. The red line is the fitting to the experimental data using Eq. \ref{eqn4} with both of the two interactions included, when force range is $\lambda = 500~\mu$m.
The initial polarization obtained by this fit is $P_0 = 4.7 \pm 0.1 \%$.
When $\lambda = 500~\mu$m, the fitting provides $g_A^eg_A^e/4\pi\hbar c =  (-0.78 \pm 1.46)\times 10^{-20}$.  The value of the axial-vector field induced interaction is less than its standard deviation showing no evidence  of the exotic interaction observed in our experiment.
The upper limit of this interaction at $\lambda = 500~\mu$m due to the statistical errors can be set to be $g_A^eg_A^e/4\pi\hbar c \leq  3.64 \times 10^{-20}$ with 95$\%$ confidence level.
The constraint due to the statistical errors can be obtained for any given force range $\lambda$ with the same procedure.

\begin{table}
 \caption{Summary of the systematic errors in our experiment. The corrections to $g_\textrm{A}^\textrm{e}g_\textrm{A}^\textrm{e}/4\pi\hbar c$ at $\lambda = 500~\mu$m are listed.}
 \label{table1}
 \begin{tabular}{l c c}
  \hline
  \hline

 Systematic error  &Size of effect & Corrections \\
  \hline
 Deviation in $x-y$ plane &  $0\pm 10~\mu$m & $(-0.6\pm 1.3)\times 10^{-20}$\\
 Distance & $12\pm1.3~\mu$m & $(1\pm 80)\times 10^{-22}$ \\
 Decoherence of $S$& $405\pm23~\mu$s &  $(-55\pm 6)\times 10^{-22} $ \\
  Decay time & $7\pm 1~\mu$s &  $(-5\pm 36)\times 10^{-21}$ \\
  Radius & $35\pm 5~\mu$m  & $(-3 \pm 7)\times 10^{-21} $ \\
  Thickness  & $15\pm3~\mu$m& $(-9\pm 45)\times 10^{-21} $\\
  Polarization &  $4.7\pm0.1\%$ &  $(-1\pm 52)\times 10^{-22}$\\
  Total & &  $(-2.9\pm 6.0)\times 10^{-20} $\\
  \hline
  \hline
\end{tabular}
\end{table}

\begin{figure}
\centering
\includegraphics[width=\columnwidth]{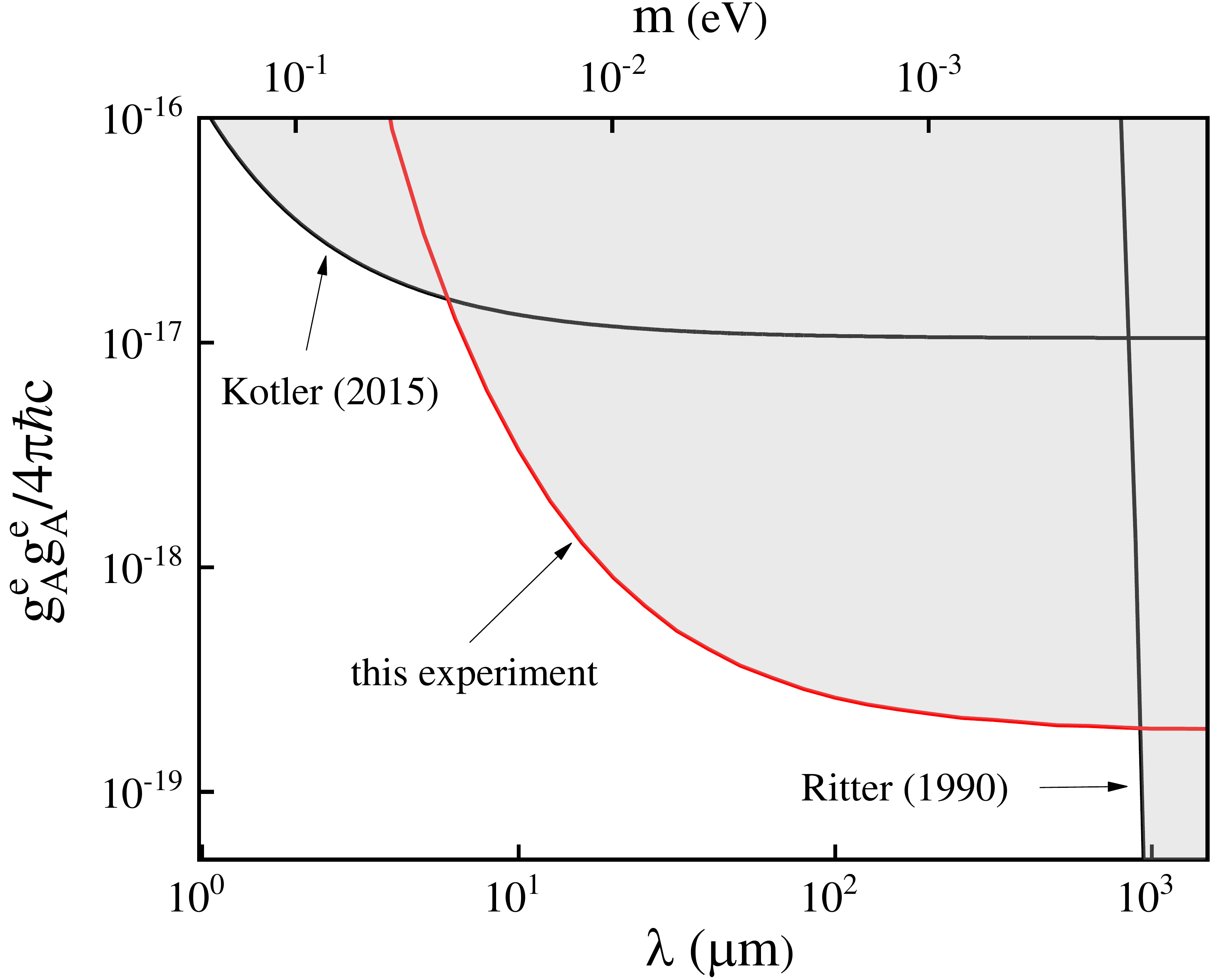}
\caption{ Upper limit on the axial-vector-mediated dipole-dipole interactions between electrons $g_\textrm{A}^\textrm{e}g_\textrm{A}^\textrm{e}/4\pi\hbar c$ as a function of the force range $\lambda$ and mass of the axial-vector bosons $m$. The black solid lines represent upper bounds from Refs. \cite{PRL_ConstraintDipDip,Ritter_PRD_1990}.  Our work (the red line) establishes a new laboratory bound in the force range from  10 to 900 $\mu$m. The obtained upper bound of the interaction at 500 $\mu$m is $|g_A^eg_A^e / 4\pi\hbar c |\leq 1.8\times 10^{-19} $, which is one order of magnitude more stringent than a previous experiment.
\label{fig3}}
\end{figure}

We examined systematic errors and analyzed the corrections to $g_A^eg_A^e/4\pi\hbar c$. We take $\lambda = 500~\mu$m as an example, while corrections due to these systematic errors are listed in Table \ref{table1}.
The main systematic error in our experiment is those of the magnetic dipole-dipole interaction between electrons due to the uncertainties of experimental parameters.
For example, the distance between $S$ and the bottom of the pentacene bulk is $12\pm 1.3~\mu$m, from which the shift of the magnetic field felt by $S$ due to the dipole-dipole interaction is estimated to be 1.0(115)$\times 10^{-10}~$T.
Then, a correction to $g_A^eg_A^e/4\pi\hbar c$ for $500~\mu$m due to this type of systematic error  can be obtained as $1(80)\times10^{-22}$.
The deviation in the $x-y$ plane is mainly due to the long time drift of our optical system, which was observed to be less than $10~\mu$m during our experiment. This effect causes a correction to the coupling of $-0.6(1.3)\times10^{-20}$.
The systematic errors due to the uncertainty of the radius, the thickness of the single crystal and the relaxation time of the polarized electron have been taken into account in the Supplemental Material \cite{SupplementalMaterial}.
The correction due to the decoherence time of $S$ is also examined. The detailed analysis of the systematic errors are included in Table \ref{table1}.
The total correction to the interaction at $500~\mu$m is $-2.9(6.0)\times10^{-20}$.
The bound for the exotic interaction with force range $\lambda = 500~\mu$m is derived to be $|g_A^eg_A^e/4\pi\hbar c|\leq 1.8\times 10^{-19}$ with a 95$\%$ confidence level, when both statistical and systematic errors are taken into account.
The upper limits  with different values of force range shown in Fig. \ref{fig3} are obtained with the same method.

Figure \ref{fig3} shows the new constraint set by this work together with recent constraints from experimental searches for axial-vector-mediated dipole-dipole interactions.
Filled areas correspond to excluded values.
For the force range $\lambda > 900 ~\mu$m, the constraint was established by Ritter $et~al$.\cite{Ritter_PRD_1990,Kimball}.
For the force range $\lambda < 10~\mu$m in Fig. \ref{fig3}, the upper limit was set by Kotler $et ~al$.\cite{PRL_ConstraintDipDip}.
The red line is the constraint established by our experimental observation, which clearly shows more stringent constraints in the range from 10 to 900$~\mu$m.
Specifically, the obtained upper limit of the exotic dipole-dipole interaction at 500~$\mu$m is about a factor of 50 more stringent than the one set by Ref. \cite{PRL_ConstraintDipDip}.
The constraint may be further improved by several strategies in the future. By enhancing the power of the excited laser, the polarization of the electron spin can be improved. Multiple laser pumping pulses can be employed together with multipulse dynamical decoupling sequence. Therefore, the accumulated phase due to the polarized electron can be enhanced. To reduce the systematic errors, one may fabricate the single crystal of  pentacene with more precision. The location of the NV center can be addressed more precisely by high resolution imaging technology, such as stimulated emission depletion microscopy \cite{NV_STED}.

\emph{Conclusion}.  We present an experimental platform to constrain an exotic dipole-dipole interaction between electrons.  Our method benefits from the high controllability of the quantum states of NV centers \cite{NatureCommun_FTGate}, which have been employed as sensitive magnetometers. Our recent work shows that NV centers can be utilized as a quantum sensor to detect the monopole-dipole interaction between an electron spin and nucleons at micrometer scale \cite{Rong_2018}.
In the present study, a new constraint on an axial-vector mediated interaction between electrons for the force range 10-900$~\mu$m  has been established. In the future, we expect that other types of spin-dependent forces \cite{JHighEnergyPhys_MacroForce} might be investigated by the NV-center  quantum sensor.  NV centers will not only be an important quantum sensor for physics within the standard model, but will also be a platform for probing hypothetical particles beyond the standard model.

This work was supported by the NSFC (Grants No. 81788101, No. 11227901, No. 11722327, No. 11653002, No. 11421303, and No. J1310021), the CAS (Grants No. GJJSTD20170001, No. QYZDY-SSW-SLH004, and No. QYZDB-SSW-SLH005), the 973 Program (Grants No. 2013CB921800 and No. 2016YFB0501603), and Anhui Initiative in Quantum Information Technologies (Grant No. AHY050000).
X.\ R and F.\ S. thank the Youth Innovation Promotion Association of Chinese Academy of Sciences for the support.
Y. F. C. is supported in part by the CAST Young Elite Scientists Sponsorship Program (2016QNRC001) and by the Fundamental Research Funds for the Central Universities.




\end{document}